\documentclass[sigconf]{aamas} 

\usepackage{times}
\usepackage{soul}
\usepackage{url}
\usepackage[utf8]{inputenc}
\usepackage{graphicx}
\usepackage{amsmath}
\usepackage{amsthm}
\usepackage{booktabs}
\usepackage{algorithm}
\usepackage{algorithmic}
\usepackage[switch]{lineno}
\usepackage[normalem]{ulem} 

\usepackage{packages/packages}
\usepackage{packages/scribe}
\usepackage{listings}
\usepackage{packages/commands}
\usepackage{packages/mymacros}
\usepackage[	
backend=biber,	
style=numeric,	
citestyle=numeric	
]{biblatex}	
\bibliography{aamas2024}

\setcopyright{ifaamas}
\acmConference[AAMAS '24]{Proc.\@ of the 23rd International Conference
on Autonomous Agents and Multiagent Systems (AAMAS 2024)}{May 6 -- 10, 2024}
{Auckland, New Zealand}{N.~Alechina, V.~Dignum, M.~Dastani, J.S.~Sichman (eds.)}
\copyrightyear{2024}
\acmYear{2024}
\acmDOI{}
\acmPrice{}
\acmISBN{}

\acmSubmissionID{999}

\title{Banzhaf Power in Hierarchical Voting Games}

\author{John Randolph}
\affiliation{
  \institution{Brown University}
  \city{Providence}
  \country{United States}
}
\email{john_randolph@alumni.brown.edu}

\author{Denizalp Goktas}
\affiliation{
  \institution{Brown University}
  \city{Providence}
  \country{United States}
}
\email{denizalp_goktas@brown.edu}

\author{Amy Greenwald}
\affiliation{
  \institution{Brown University}
  \city{Providence}
  \country{United States}
}
\email{amy_greenwald@brown.edu}

\begin{abstract}
The Banzhaf Power Index (\bpi) is a method of 
measuring the power of voters in determining the outcome of a voting game.
Some voting games exhibit a hierarchical structure, including the US electoral college and ensemble learning methods; we call such games hierarchical voting games.
It is generally understood that \bpi{} in hierarchical voting games can be computed via a recursive decomposition of the hierarchy, which can substantially reduce the calculation's complexity.
We identify a key (previously undocumented) assumption on which this decomposition is based, namely balance, meaning one group of voters has enough votes to win whenever the complementary group of voters does not, and vice versa.
We then introduce a generalization of BPI that we call Extended \bpi{} (\ebpi) for all voting games, including those that are not balanced, which simplifies to \bpi{} in balanced games.
We show that \bpi{} in unbalanced hierarchical voting games decomposes in terms of \ebpi{} at each level in the hierarchy, which yields computational savings analogous to those achieved in the balanced case.
As a sample application, we take advantage of the compositionality of language, and model the impact of individual words on a sentence's sentiment as a voting game.
As the complement of a phrase in a sentence does not necessarily have the opposite sentiment, this voting game is unbalanced and requires our decomposition of \bpi{} in terms of \ebpi{}.
Our results suggest that EBPI is an effective proxy for BPI
(because the meaning of a sentence is not always 100\% compositional),
and demonstrate a dramatic improvement in run time.
\end{abstract}

\keywords{Banzhaf Power Index, Cooperative Game Theory, Voting Games, Social Choice}

\begin{document}


\pagestyle{fancy}
\fancyhead{}


\maketitle 


\section{Introduction}

Elections are a form of group decision-making intended to produce outcomes that reflect the collective preferences of the group, or at least give the impression of doing so.
The designers of an election
might want their system to embody principles of fairness, such as ``one person, one vote'' \cite{auerbach1964,goel2020}.
In complex elections, it is often unclear how much influence an individual voter has on the outcome, making it difficult to assess whether the system is fair or not.
Measures of power can help us determine whether our elections meet our standards of fairness.

Elections can be modeled 
as a type of cooperative game called voting games~\cite{hsaio1993,young2015}.
A voting game comprises a set of voters and a characteristic function, the latter of which takes as input a coalition, i.e., a subset of voters, and returns a binary variable.
If the characteristic function outputs a 1 (resp.\ 0), we say that the coalition is a winning (resp.\ losing) coalition.

For example, if we model the United States (US) electoral college as a voting game, then the set of voters is the set of states, and the characteristic function returns 1 whenever the number of electoral votes associated with a coalition, i.e., a set of states, sums to at least 270.
Thus, any coalition whose total number of electoral votes is at least 270 is a winning coalition, and every other is a losing coalition.

Electoral processes in representative democracies often have hierarchical structure.
In the US, for example, voters vote for electoral college delegates,
who then vote to elect the president.
Likewise, voters in Slovenian National Council elections vote for a representative in their local councils, who in turn vote for national councillors, who in turn vote on legislation in the
National Council.
We call these voting games hierarchical, as they are structured as trees, with
voters at the leaf nodes, such that votes propagate up the tree to intermediate nodes, with the final result collectively determined by characteristic functions at each level.

Another example of a hierarchical voting game is an ensemble learning model, which combines the results of multiple learning algorithms in a hierarchical manner in attempt to obtain better performance than any of the algorithms alone \cite{Kulynych2017features}.
When attempting to interpret an ensemble's predictions, it can be instructive to understand the impact of each individual algorithm on the ensemble's overall prediction \cite{karczmarz2021tree}.

In 1965, Banzhaf defended a mathematical measure of power in voting games \cite{Banzhaf1965weighted} first introduced by Penrose \cite{penrose1946} but then largely forgotten \cite{felsenthal1998}, now called the Banzhaf power index (\bpi).
\bpi{} has since found many applications, such as the US electoral college \cite{Barthelemy2011apportionment}, the various councils and parliaments within the European Union \cite{algaba2007constitution,Ferto2020agricultural,Kirsch2016Brexit,Rapoport1985israel,SeijasMarcias2019finnish}, the International Monetary Fund \cite{Aleskerov2008imf}, feature importance in machine learning \cite{Kulynych2017features}, and shareholders and corporate boards \cite{Kirstein2010volkswagen,Leech1988shareholding}.

The formula for computing \bpi{} is exponential:
each of the $2^\numplayers$ coalitions must be examined to determine whether it is a winning coalition, with and without each of the $\numplayers$ voters.
This leads to a time complexity of $O(2^\numplayers)$ per voter
\cite{klinz2005faster}, which can be prohibitive, for example, when $\numplayers$ is the number of eligible voters in the US.

\bpi{} in hierarchical voting games, however, decomposes in terms of \bpi{} at each of the $d$ levels in the hierarchy, according to a formula we call multiplicative \bpi{} (\mbpi) \cite{miller2013}.
This decomposition provides a massive speedup when the branching factor $\branchingfactor \ll \numplayers$:
it only takes $O(\depth 2^\branchingfactor)$ time per voter.
Our first contribution is to identify the key mathematical property of a voting game that is necessary for this decomposition, namely balance, i.e., if a coalition of voters does not have enough votes to win then its complement does and vice versa.

It is well known that \bpi{} ascribes more power to voters who reside in large states than in small states in US presidential elections \cite{diss2021}.
Indeed, Banzhaf's analysis of the 1964 electoral college found that a voter in California had 3.312 times as much power as a voter in the District of Colombia \cite{banzhaf1968}.
This decomposition theorem allows us to understand why.
A voter's \bpi{} is a product of their influence on their state's election and their state's influence on the national election.
Thus, although an individual voter in California has less influence on how California votes than a voter in DC has on how DC votes, California has \emph{much\/} more influence on the national election than DC, because of its large number of electoral votes.

Although many common voting games are balanced, others are not.
For example, if a supermajority (not a simple majority) is required to win, then the corresponding voting game is not balanced.
In the US Senate, for example, 60 votes are required to pass most legislation; as a result, every set of 41 to 59 senators is a losing coalition, as is its complement.
Perhaps surprisingly, it turns out that the aforementioned \bpi{} decomposition theorem does not hold in unbalanced hierarchical games.

\paragraph{Contributions}
In search of a similar decomposition theorem for unbalanced hierarchical voting games, we introduce a generalization of \bpi{} that we call extended \bpi{} (\ebpi), which is applicable to all monotone 
voting games, balanced or not.
Then, just as \bpi{} can be calculated efficiently via \mbpi{} in balanced hierarchical games, we propose multiplicative \ebpi{} (\embpi) to efficiently calculate \ebpi{} in unbalanced hierarchical games.
Our main theorem states that \embpi{} in fact yields a novel, more efficient means of calculating \bpi{} for this larger class of games.

The \ebpi{} formula, which generalizes \bpi, is a factor of $\branchingfactor$ slower than the naive \bpi{} formula.
Furthermore, the \embpi{} formula depends on the number of winning and losing coalitions at \emph{every\/} subgame in the hierarchy, and thus involves a full tree traversal.
As a result, calculating \embpi{} is substantially slower than calculating \mbpi, so should only be used when necessary, namely in unbalanced games.
But like \mbpi{} for balanced games, \embpi{} for unbalanced games is a vast improvement over the naive approach.

We summarize the complexity of our formulas as compared to naively computing \bpi{} in \Cref{table:runtimes}.
Methods to approximate \bpi{} via random sampling have also been proposed, with 
computational complexity that is polynomial in accuracy and confidence level, but they cannot achieve accuracy better than $\nicefrac{1}{2^{\numplayers^\epsilon}}$ for $\epsilon > 0$, without a number of samples that is exponential in $\numplayers$~\cite{bachrach2009approximating}.

\begin{table}[t]
    \centering
    \begin{tabular}{llll}
        \toprule
        & \text{Naive}    & \text{Balanced Tree}    & \text{General Tree} \\
        \midrule
        \text{One voter} & $O(2^\numplayers)$ & $O(\depth 2^\branchingfactor)$ & $O(\numplayers \depth \branchingfactor 2^\branchingfactor)$ \\
        \text{All voters} & $O(\numplayers 2^\numplayers)$ & $O(\numplayers \depth 2^\branchingfactor)$ & $O(\numplayers \depth \branchingfactor 2^\branchingfactor)$ \\
        \bottomrule
    \end{tabular}
\caption{\bpi{} algorithm run times, where $\depth$ is the depth of the tree, $\numplayers$ is the number of voters, and $\branchingfactor$ is the branching factor.}
\label{table:runtimes}
\end{table}

To demonstrate the utility of our algorithms, we calculate the \bpi{} of individual voters in Velenje and Ljubljana in Slovenian National Council elections.
We find that the \bpi{} of a voter in Velenje is about three times that of a voter in Ljubljana.
Note that it would be intractable to calculate these values taking a naive approach, as it would involve examining all possible subsets of the 2 million Slovenian voters.

We then turn our attention to the problem of vocabulary selection in sentiment analysis, which we tackle by finding the words that are most influential 
(as measured by approximate BPI) in determining a text's sentiment~\cite{patel2021}.
Because the structure of grammar is inherently hierarchical, we model word importance as a hierarchical voting game.
Our approach is only approximate, however, because sentiment is not perfectly compositional.
Thus, we obtain only an approximation of \bpi, albeit one that can be calculated much faster than the exact value.
We empricially compare our approximate values to exact BPI on small problem instances, and conclude that our approximations are acceptable.

An alternative popular power index is \mydef{Shapley-Shubik's} (\mydef{SSPI}) \cite{shapley1954}.
SSPI, which involves enumerating every permutation of voters, of which there are $\numplayers$!, has been used to model power distributions in a broad range of applications, including the US electoral college \cite{wright2009}, feature importance in machine learning \cite{karczmarz2021tree}, and shareholders on corporate boards \cite{prigge2007}.
Although this paper is not concerned with SSPI, the techniques we develop in this paper may provide directions for future research on the computation of SSPI.

This paper is organized as follows.
First, we introduce notation surrounding voting games and power indices.
Then, we explain how to compute \bpi{} in balanced hierarchical games, with an example application, namely the Slovenian National Council.
Next, we introduce our generalized power measure (\ebpi), as well as our algorithm (\embpi) to compute it on all hierarchical voting games, balanced or not, and show that it in fact computes \bpi.
Finally, we run an experiment to demonstrate our approach on vocabulary selection.

\section{Voting Games}

A \mydef{voting game}, also called a simple game or a simple voting game, $\game \doteq (\players, \charfn)$ consists of a set of $\numplayers$ voters $\players$, who can form \mydef{coalitions} (i.e., subsets) that can either win or lose an election according to a binary characteristic function $\charfn: 2^{\players} \rightarrow \{0, 1\}$.

The coalition $\coalition \subseteq \players$ is a \mydef{winning coalition} if $\charfn(\coalition) = 1$, and a \mydef{losing coalition} if $\charfn(\coalition) = 0$.
The \mydef{complement} of a coalition $\coalition \subseteq \players$ is the coalition $\players \setminus \coalition$.
The number of winning (resp.\@ losing) coalitions is given by
$| \{\coalition \subseteq \players \mid \charfn (\coalition) = 1 \} |$
(resp.\@ $| \{\coalition \subseteq \players \mid \charfn (\coalition) = 0 \} |$).

A game is said to be \mydef{balanced}, or proper strong, if the complement of every winning (resp.\ losing) coalition is a losing (resp.\ winning) coalition.
That is, given $\coalition \subseteq \players$ in a balanced voting game $\game = (\players, \charfn)$, $\charfn(\coalition) = 1$ if and only if $\charfn(\players \setminus \coalition) = 0$.
Consequently, there are the same number of winning and losing coalitions in a balanced game, namely $2^{n-1}$.

In this paper, we restrict attention to voting games with \mydef{monotone} characteristic functions: i.e., for all coalitions $\coalition \subseteq \players$ and voter $\player \in \players \setminus \coalition$, $\charfn(\coalition \cup \{ \player \}) = 1$ whenever $\charfn(\coalition) = 1$.
In other words, adding a voter to a winning coalition cannot turn it into a losing coalition.
Furthermore, $\charfn(\emptyset) = 0$ and $\charfn(\players) = 1$.

A \mydef{weighted voting game} $(\players, \weight, \quota)$ with voters $\players$, weight vector $\weight \in \mathbb{N}_+^{\numplayers}$, and quota $\quota[ ] \in \mathbb{N}_+$ is a voting game $\game = (\players, \charfn)$ with \mydef{quota-based} characteristic function
\begin{align}
    \charfn(\subsetplayers) = \left\{
    \begin{array}{cl}
        1 & \text{if} \sum\limits_{\player \in \subsetplayers} w_\player \geq \quota[ ] \\
        0 & \text{otherwise}
    \end{array} \right.
    \enspace ,
\label{eq:weighted-char-fn}
\end{align}

\noindent
One example of a weighted voting game is the US Electoral College, where the voters are the states, the weights are their electoral votes, and the quota is 270.

A \mydef{weighted majority (voting) game} $(\players, \weight)$ is a weighted voting game where $\quota[ ] = \frac{1}{2}\sum\limits_{\player \in \players} \weight[i]$.

\subsection{Hierarchical Voting Games}
We begin by describing the syntax of hierarchical voting games.
We then introduce some standard tree nomenclature in the context of hierarchical voting games, which we use to explain their semantics.

\paragraph{Syntax}
We call a voting game $\game$ \mydef{hierarchical} if it has the structure of a tree $(\players, \nodes, \{ \edges[\node] \}_{\node \in \nodes \setminus \players}, \{ \charfn_\node \}_{\node \in \nodes \setminus \players})$
with nodes $\nodes$, of which voters $\players \subseteq \nodes$ are leaf nodes.
Each node $\node \in \nodes \setminus \players$ has a distinct set of children $\edges[\node] \subseteq \nodes$, and is characterized by a local characteristic function $\charfn_\node : 2^{\edges[\node]} \rightarrow \{0, 1\}$, which denotes whether a coalition $\coalition \subseteq \edges[\node]$ wins or loses a local voting game.
The \mydef{branching factor} of such a voting game is $\branchingfactor = \max_{\node \in \nodes \setminus \players} |\children[\node]|$.

A \mydef{hierarchical weighted voting game} $\game =  
(\tree, \weight, \quota)$ is a hierarchical voting game $\tree$ together with local characteristic functions defined recursively, given a weight vector $\weight \in \mathbb{N}_+^{|\nodes|}$ and a quota $\quota[\node] \in \mathbb{N}_+$, as follows: $\charfn_\node (\subsetplayers) = \mathds{1} \left( \sum \limits_{\player \in \edges[\node]} \charfn_\player (\subsetplayers) \weight[\player] \geq \quota[\node] \right)$, for all nodes $\node \in \nodes \setminus \players$ and $\subsetplayers \subseteq \players$;
and $\charfn_\player (\subsetplayers) = \mathds{1} (\player \in \subsetplayers)$, for all $\player \in \players$.%
\footnote{Here, $\mathds{1}$ denotes an indicator function, which takes value $1$ whenever the parenthetical condition holds, and $0$ otherwise.}
The characteristic function at a leaf node $\player$ simply returns whether $\player \in \subsetplayers$; the characteristic functions at non-leaf nodes are quota-dependent.
A \mydef{hierarchical weighted majority (voting) game} is a hierarchical weighted voting game in which $\quota[\node] = \frac{1}{2}\sum\limits_{\player \in \edges[\node]} \weight[\player]$, for all nodes $\node \in \nodes \setminus \players$.
We denote a hierarchical weighted majority game by tuples $\game = (\tree, \weight)$.

\paragraph{Tree Nomenclature}
Before we can describe the semantics of hierarchical voting games, we introduce some standard tree nomenclature in the context of hierarchical voting games.

Given a hierarchical voting game $\game$, as the set $\edges[\node]$ denotes the \mydef{children} 
of a node $\node \in \nodes$, $\node$ is called a \mydef{parent} of each node $\player \in \edges[\node]$.
The tree structure ensures that all nodes $\node$ have exactly one parent $\parent{\node}$, except for the \mydef{root}, denoted $\treeroot$, which does not have a parent.
The \mydef{ascendants} of $\node$, denoted $\ascendants[\node] \subseteq \nodes$, is the set of nodes
\begin{align*}
\ascendants[\node] =  
\left\{ 
\begin{array}{cl}
    \emptyset & \text{if $\node$ is the root} \\
    \parent{\node} \cup \ascendants[\parent{\node}]
    & \text{otherwise} \\
\end{array}
\right.
\end{align*}
The \mydef{descendants} of $\node$, denoted $\descendants[\node] \subseteq \nodes$, is the set of nodes
\begin{align*}
\descendants[\node] =  
\left\{ 
\begin{array}{cl}
    \emptyset & \text{if $\node$ is a leaf} \\
    \children[\node] \bigcup \{ \descendants (\playertwo) \mid \playertwo \in \children[\node] \}
    & \text{otherwise} \\
\end{array}
\right.
\end{align*}

\noindent

Given a node $\node \in \nodes$ in a hierarchical voting game $\tree$,
we write $\leaf[\node]  = \players \cap \descendants[\node]$ to denote all the leaf nodes that descend from $\node$.
Then, the \mydef{subgame} $\subgame{\game}{\node} = (\leaf[\node], \{ \node \} \cup \descendants[\node], \{ \edges[\node] \}_{\node \in \descendants[\node]}, \{ \charfn_\node \}_{\node \in \descendants[\node]})$ is the hierarchical voting game with $\node$ as the root.
The \mydef{local subgame} $\onesubgame{\node}$ is the (flat) voting game $(\edges[\node], \charfn_\node)$.

The number of winning and losing coalitions in a hierarchical voting game is the number of winning (resp.\@ losing) coalitions that can be formed by the voters in the game.
More formally, the number of winning coalitions in subgame $\subgame{\game}{\node}$, with root note $\node$ and implicit characteristic function $\charfn$,
is given by
\begin{align*}
\numwin[\node] =  
\left\{ 
\begin{array}{cl}
    1 & \text{if $\node$ is a leaf} \\
    | \{\coalition \subseteq \leaf[\node] \mid \charfn (\coalition) = 1 \} | & \text{otherwise} \\
\end{array}
\right.
\end{align*}
The number of losing coalitions is defined analogously.
\noindent
Likewise, the number of losing coalitions is given by
\begin{align*}
\numlose[\node] =  
\left\{ 
\begin{array}{cl}
    1 & \text{if $\node$ is a leaf} \\
    | \{\coalition \subseteq \leaf[\node] \mid \charfn (\coalition) = 0 \} | & \text{otherwise} \\
\end{array}
\right.
\end{align*}

\paragraph{Semantics}
We can now return to the semantics of hierarchical voting games.
We sometimes refer to non-hierarchical voting games as ``flat;'' such games can also be understood as hierarchical games of depth $1$, where all the voters are siblings: i.e., children of the root.

Every hierarchical voting game represents a flat voting game, with an \mydef{implicit} characteristic function $\charfn$ defined recursively as follows:
for $\subsetplayers \subseteq \players$, $\charfn (\subsetplayers) =  
\charfn^\prime (\subsetplayers; \treeroot)$, where $\charfn^\prime (\subsetplayers; \node) = \charfn_\node ( \{ \player \in \edges[\playerthree] \mid \charfn^\prime (\subsetplayers; \player) = 1 \})$, for all $\node \in \nodes \setminus \players$;
and $\charfn^\prime (\subsetplayers; \player) = \mathds{1} (\player \in \subsetplayers)$, for all $\player \in \players$.%
\footnote{By monotonicity, it must be the case that, at a leaf node $\player \in \players$, there is always exactly one way to win (the coalition $\{ \player \}$ itself) and one way to lose (the empty set).}
In words, $\charfn^\prime (\subsetplayers; \node)$ represents whether or not $\subsetplayers$ is a winning coalition in the subgame $\subgame{\game}{\node}$.
This implicit characteristic function is constructed by applying the local characteristic function $\charfn_\node$ to all of $\node$'s children $\player \in \edges[\node]$ for which $\subsetplayers$ is a winning coalition in $\subgame{\game}{\player}$:
i.e., local characteristic functions are applied recursively from node $\node$ down to the leaves.

A large class of weighted majority games, hierarchical or otherwise, namely ones for which ties cannot occur, are balanced,
because each winning coalition has a majority of the weight, so its complement has a minority of the weight and is thus a losing coalition.%
\footnote{See the supplementary material for a technical specification of these games.}

\section{Banzhaf Power Index}

A \mydef{power index} is a function 
$\pow$, which, given a voting game, associates a real value with each voter, which we interpret as
power.
The \mydef{Banzhaf power index (\bpi)} is one such index.
\bpi{} for a voter $\player$ is calculated by examining all subsets of voters that exclude voter $\player$.
More precisely, \bpi{} equals the number of such subsets that is a winning coalition with $\player$ and a losing coalition without $\player$, divided by the total number of such subsets.
For example, in a game with three voters, there are four coalitions that do not include voter $\player$.
If three among them are losing coalitions without voter $\player$, which become winning coalitions with voter $\player$, then the \bpi{} of voter $\player$ is $\frac{3}{4}$.

Mathematically, given a voting game $\game$,
\begin{align}
\label{eq:bpi}
    \bhpow[\player](\game) &= \frac{\sum\limits_{\subsetplayers \subseteq \players \setminus \{\player \}}(\charfn(\subsetplayers \cup \{ \player\}) - \charfn(\subsetplayers))}{2^{|\players|-1}}.
\end{align}

\noindent
As this formula incorporates the characteristic function of all voter subsets, its time complexity is $O(2^\numplayers)$ for one voter, and $O(\numplayers 2^\numplayers)$ for all voters.

When all voters in a weighted majority game have the same weight, it makes intuitive sense that they would all have the same voting power.
This property
indeed holds of \bpi.
In such games, any coalition comprising at least ``half'' ($1 + \lfloor \frac{|\players| - 1}{2} \rfloor$) the voters is a winning coalition, while any coalition comprising less than ``half'' ($\lfloor \frac{|\players| - 1}{2} \rfloor$) the voters is a losing coalition.
The following lemma is well-known:%
\footnote{All proofs appear in the supplementary material.}

\begin{restatable}
{lemma}{lemmaequalpower}
Given a 
weighted majority game $\game = (\players, \weight, \quota[ ])$, if all voters $\player \ne \playertwo \in \players$ have equal weight, then $\bhpow[\player](\game) = \bhpow[\playertwo](\game) = \binom{|\players| - 1}{\lfloor \frac{|\players| - 1}{2} \rfloor}
\left( \frac{1}{2^{|\players| - 1}} \right)$.
\label{lemma:equal-power}
\end{restatable}

Given a hierarchical voting game $\game = (\players, \nodes, \{ \edges[\node] \}_{\node \in \nodes}, \{ \charfn_\node \}_{\node \in \nodes})$, we define the \mydef{multiplicative Banzhaf power index (\mbpi)} of voter $\player \in \players$ in game $\game$ as follows:
\begin{align*}
\multbhpow[\player] (\game) =  
\left\{ 
\begin{array}{cl}
    1 & \text{if $\player$ is the root} \\
    \prod\limits_{\playertwo \in \ascendants[\player] \cup \{ \player \}} \bhpow[\playertwo] (\onesubgame{\parent{\playertwo}}) & \text{otherwise} \\
\end{array}
\right.
\end{align*}

\noindent
In other words, the \mbpi{} of a voter $\player$ is the product of \bpi{} at each of $\player$'s ancestors.

\mbpi{} is of interest because it recovers \bpi{} in \emph{balanced} hierarchical voting games, where the complement of every winning (resp.\ losing) coalition is a losing (resp.\ winning) coalition.
This makes sense intuitively: the power of a voter (represented as a leaf in the tree) equals their voting power in their local election, multiplied by the voting power of their representative (represented by their parent) in their local election, and so on.
That the game is balanced ensures that the computation only involves the path from the root to the voter, as no area of the hierarchy can command outsized influence.

\begin{restatable}{theorem}{thmbalancedequal}
\label{thm:balanced-equal}
    In balanced hierarchical voting games, the multiplicative Banzhaf power index is equal to the Banzhaf power index: i.e.,  $\multbhpow[\player](\game) = \bhpow[\player](\game)$, for all $\player \in \players$.
\end{restatable}

The \mbpi{} formula gives rise to a more computationally efficient way to compute \bpi{} for a given voter $\player$.
Rather than enumerating all $O(2^{\numplayers})$ voter subsets, only $O(\depth 2^{\branchingfactor})$  enumerations are required, one $O(2^{\branchingfactor})$ enumeration at each of the $\depth$ nodes along the path to voter $\player$.
As $\branchingfactor \leq \numplayers$, calculating \mbpi{} is always at least as fast as \bpi, and often much faster.
The smaller the branching factor, the greater the speed-up.

\paragraph{Slovenian national council}

The US electoral college is a small enough weighted majority game for us to calculate \bpi{} directly for all 50 states
\cite{banzhaf1968,chandler2022}.
This process yields a measure of power for each state \cite{Barthelemy2011apportionment}.
But many other systems are too large and complicated for us to compute \bpi{} directly.

Slovenia's national council is elected via a complex process: there are 40 indirectly elected members, 22 that represent municipalities and 18 that represent special interests, such as sports and culture, or farmers.
Each of these members is elected by an electoral college: for the 22 members that represent municipalities, these electoral colleges are the local assembly, and for the other 18 members, these electoral colleges include members of the sector they represent \cite{senat2023}.
As such, this voting game is hierarchical.

Calculating the power of any individual Slovenian voter on a piece of legislation in the Slovenian National Council by computing \bpi{} naively would be far too expensive, since it would involve examining all $2^\numplayers$ coalitions, where $\numplayers$ is the number of voters in Slovenia (around 2 million).

Note also that we cannot just invoke \Cref{lemma:equal-power}, because although individual voters have equal weight within their municipalities in electing their local assembly, the hierarchy obscures their weight relative to other voters in Slovenia.
Still, within each local subgame (voters voting in their local assembly; local councillors voting for national councillors in their electoral college; national councillors voting on legislation) we \emph{can\/} make use of \Cref{lemma:equal-power}, as there, voters have equal weights.
\sjohn{}{}
Then, since the voting system belongs to the class of weighted majority games without ties%
\footnote{Note that ties do not occur in the passing of legislation in this case, because in the event of a 20-20 split of the council, the group that voted for the legislation loses and the group that voted against it wins, which serves as an arbitrary tie-breaking mechanism 
\cite{jambrek1990}. In many other parliamentary elections, there is a tie-breaking voter: e.g., in the U.S. Senate, where the vice president breaks ties in the case of a 50-50 split.}
and thus is necessarily balanced, we can compute \bpi{} via \mbpi.
In other words, the power of a voter in the Slovenian National Council equals the product of the power of that voter in electing their representative, the power of their local councillor in their municipality's electoral college, and the power of their municipality's national councillor in the National Council.

Now we can ask the question: who has more power in the Slovenian National Council, a voter in the small municipality of Velenje or the large municipality of Ljubljana?
Table~\ref{tab:slovenian-council} summarizes the requisite information about the election of the Velenje and Ljubljana councillors to answer this question \cite{senat2023,ljubljana2023,velenje2023}.

\begin{table}[t]
    \centering
    \begin{tabular}{lll}
        \toprule
        \text{Municipality} & \text{\footnotesize Voting population} 
        & \text{\footnotesize Local council size}\\
        \midrule
        \text{Velenje} & 10,039 & 33 \\
        \text{Ljubljana} & 65,041 & 45 \\
        \bottomrule
    \end{tabular}
\caption{Slovenian National Council Election Data 2023.}
\label{tab:slovenian-council}
\end{table}

Take Velenje as an example.
The power of a voter in Velenje in electing their local councillor is $\frac{\binom{|\players| - 1}{\lfloor \frac{|\players| - 1}{2} \rfloor}}{ 2^{|\players| - 1}} = \frac{\binom{10038}{5019}}{ 2^{10,038}} = 0.00796$;
the power of their local councillor in choosing Velenje's national councillor is $\frac{\binom{|\players| - 1}{\lfloor \frac{|\players| - 1}{2} \rfloor}}{ 2^{|\players| - 1}} = \frac{\binom{32}{16}}{2^{32}} = 0.13995$; 
and the power of the national councillor in the National Council is $\frac{\binom{|\players| - 1}{\lfloor \frac{|\players| - 1}{2} \rfloor}}{ 2^{|\players| - 1}} = \frac{\binom{39}{19}}{2^{39}} = 0.12537$.
As per \mbpi, the power in the National Council of an individual voter in Velenje is the product of these three values: $(0.00796)(0.13995)(0.12537) = 0.00013966$.

In Table~\ref{tab:slovenian-council-results}, we summarize the results from this calculation for a Velenje voter and a Ljubljana voter.
We find that a voter in Velenje has about 3 times the power of a voter in Ljubljana.
This finding challenges the well-known big-state advantage in the US, in which voters who reside in more populous states are attributed more power via \bpi{} \cite{Banzhaf1965weighted}.
A possible explanation is that the local council size is not linear in the size of the voting population; indeed, Velenje seems very well represented relating to Ljubljana.

\begin{table}[t]
    \centering
    \begin{tabular}{lll}
        \toprule
        \text{Municipality}    & \text{\footnotesize P.\ voter in councillor} & \text{\footnotesize P.\ councillor in E.C.} \\
        \midrule
        \text{Velenje} & $0.00796$ & $0.13995$ \\
        \text{Ljubljana} & $0.00312858$ & $0.119604$ \\
        \bottomrule
        \toprule
        \text{Municipality} & \text{\footnotesize P.\ councillor in N.C.} & \text{\footnotesize P.\ voter in N.C.} \\
        \midrule
        \text{Velenje} & $0.12537$ & $0.00013966$ \\
        \text{Ljubljana} & $0.12537$  & $0.00004691$\\
        \bottomrule
    \end{tabular}
\caption{The power of voters in the Slovenian National Council. A Velenje voter has about 3 times the power of a Ljubljana voter, as $\bm{0.00013966 / 0.00004691 \approx 2.977}$.}
\label{tab:slovenian-council-results}
\end{table}

\section{Extended Banzhaf Power Index}

Not all weighted voting games are balanced; likewise, not all hierarchical weighted voting games are balanced. 
For example, consider the US Senate: the voters in each state vote for their Senators, who then vote on legislation, with 60 of the 100 Senators' votes needed to pass most legislation.
Suppose only 55/100 Senators vote for a certain piece of legislation.
This voting game is unbalanced, because the voters who elected those 55 Senators do not comprise a winning coalition, and neither do the complementary group of voters who elected the other 45 Senators.

In this section, we introduce a novel formula to calculate \bpi{} in hierarchical voting games that are not necessarily balanced.
Like \mbpi, our more general formula takes advantage of hierarchical structure when it is present in a voting game, multiplying a voter's power in a local subgame by those of all its representatives (i.e., ancestors) in their local subgames.
The only difference is that we rely on an
\mydef{extended Banzhaf power index}, or \ebpi, which we define, to account for any lack of balance in the game.

Given a voting game $\game = (\players, \charfn)$, together with a per-voter vector of $\numwin[]$ and $\numlose[]$ values, we define the \mydef{extended Banzhaf power index (\ebpi)} of a node $\player \in \nodes$ as the natural extension of \bpi{} to voting games that are not necessarily balanced:
\begin{align}
\label{eq:ebpi}
\begin{split}
    \ebhpow[\player] (\game) & = 
\frac{\sum \limits_{\subsetplayers \subseteq \children[\parent{\player}] \setminus \{ \player \}} \Bigl[ (\charfn(\subsetplayers \cup \{ \player \}) - \charfn(\subsetplayers)) }{2^{\numplayers-1}} \\
& \frac{\prod_{\playertwo \in \children[\parent{\player}] \setminus \{ \player \}} (\mathds{1} (\playertwo \in \subsetplayers) \numwin[\playertwo] + \mathds{1}(\playertwo \notin \subsetplayers) \numlose[\playertwo]) \Bigr] }{\prod_{\playertwo \in \players \setminus \{ \player \}} 2^{|\leaf[\playertwo]|-1}} \\
\end{split}
\end{align}

Like \bpi, a voter $\player$'s \ebpi{} is a tally of the total number of coalitions to which $\player$ is critical: i.e., that are winning with $\player$ and losing without $\player$.
This number is the sum over all coalitions $\subsetplayers \subseteq \players \setminus \{ \player \}$ to which $\player$ is critical of the number of ways to combine the relevant winning and losing coalitions of $\player$'s siblings $\playertwo \in \players \setminus \{ \player \}$.
More specifically, given that $\player$ makes $\subsetplayers$, a losing coalition, into a winning coalition, we factor in the number of winning coalitions, for all $\playertwo \in \subsetplayers$, as these coalitions are still winning, and the number of losing coalitions, for all $\playertwo \not \in \subsetplayers$, as these coalitions are now winning.
The product of all of these numbers across all $\playertwo \in \players \setminus \{ \player \}$ is the total number of ways to combine the relevant winning and losing coalitions, given $\subsetplayers$, and thus the total number of coalitions to which $\player$ is critical.

In flat voting games, where $\numwin[\player] = \numlose[\player] = |\leaf[\player]| = 1$, for all voters $\player \in \players$,
\ebpi{} equals \bpi.
In other words, \ebpi's additional accounting is not necessary in flat voting games.

\begin{restatable}{lemma}{lemmaebpiequalflat}
Given a (flat) voting game $\game = (\players, \charfn)$, $\ebhpow[\player](\game) = \bhpow[\player](\game)$, for all $\player \in \players$.
\label{lemma:ebpi-equal-flat}
\end{restatable}

If a flat voting game is balanced, then there are $2^{\numplayers-1}$ winning coalitions and $2^{\numplayers-1}$ losing coalitions, as there is a one-to-one mapping between each winning and losing coalition.
More generally, if a hierarchical voting game is balanced, then the number of winning and losing coalitions in each subgame $\subgame{\game}{\node}$ is $2^{|\leaf[\node]|-1}$,
for all nodes $\node \in \nodes \setminus \players$.
Therefore, whenever a hierarchical game is balanced, all the $\numwin[\playertwo]$, $\numlose[\playertwo]$, and $2^{|\leaf[\playertwo]|-1}$ factors on the RHS of Equation~\ref{eq:ebpi} cancel, and the formula for \ebpi{} reduces to the formula for \bpi:

\begin{restatable}{lemma}{lemmaebpiequalbalanced}
If a hierarchical voting game is balanced, then $\ebhpow[\player](\game) = \bhpow[\player](\game)$, for all $\player \in \nodes$.
\label{lemma:ebpi-equal-balanced}
\end{restatable}

\noindent
In other words, \ebpi's additional accounting is ``correct,'' in 
that it yields \bpi{} in balanced hierarchical voting games.

The \mydef{multiplicative extended Banzhaf power index (\embpi)} of voter $\player \in \nodes$ in a hierarchical voting game is identical to \mbpi, except that it makes use of \ebpi{} rather than \bpi:
\begin{align*}
\emultbhpow[\player] (\game) =  
\left\{ 
\begin{array}{cl}
    1 & \text{if $\player$ is the root} \\
    \prod_{\playertwo \in \ascendants[\player] \cup \{ \player \}} \ebhpow[\playertwo] (\onesubgame{\parent{\playertwo}}) & \text{otherwise} \\
\end{array}
\right.
\end{align*}

Since \ebpi{} equals \bpi{} in balanced games, and \mbpi{} (the product of \bpi{} s across levels) equals \bpi{} in balanced hierarchical games, it follows that \embpi{} (the product of \ebpi{} s across levels) also equals \bpi{} in balanced hierarchical games: i.e., $\emultbhpow[\player](\game) = \bhpow[\player](\game)$, for all $\player \in \players$.

The next theorem states that \embpi{} equals \bpi{} in hierarchical voting games, balanced or otherwise.
The intuition for this result is the same as for \mbpi, except for the additional terms, which account for winning and losing coalitions in distant subgames, which can affect \bpi{} when the game is unbalanced, because it affords them outsized influence.

\begin{restatable}{theorem}{thmgeneralequal}
In 
hierarchical voting games, balanced or otherwise, the multiplicative extended Banzhaf power index equals the Banzhaf power index: i.e., $\emultbhpow[\player](\game) = \bhpow[\player](\game)$, for all $\player \in \players.$
\label{thm:general-equal}
\end{restatable}

To compute $\numwin[\node]$ and $\numlose[\node]$, for all nodes $\node \in \nodes$, in a hierarchical voting game, it suffices to traverse the tree and count the number of winning and losing coalitions in each subgame.
This counting can be accomplished by the following recursive formula:
\begin{align}
\numwin[\node] =  
\left\{ 
\begin{array}{cl}
    1 & \text{if $\node \in \players$} \\
    \sum \limits_{\subsetplayers \subseteq \edges[\parent{\node}]} \Bigl[ \charfn_\node(\subsetplayers) \prod\limits_{\playertwo \in \edges[\parent{\node}] \setminus \{\node \}} & \\
    (\mathds{1}(\playertwo \in \subsetplayers) \numwin[\playertwo] + \mathds{1}(\playertwo \notin \subsetplayers) \numlose[\playertwo]) \Bigr] & \text{otherwise} \\
\end{array}
\right. 
\label{eq:numwin}
\end{align}
\begin{align}
\numlose[\node] =  
\left\{ 
\begin{array}{cl}
    1 & \text{if $\node \in \players$} \\
    \sum \limits_{\subsetplayers \subseteq \edges[\parent{\node}]} \Bigl[ (1 - \charfn_\node(\subsetplayers)) \prod\limits_{\playertwo \in \edges[\parent{\node}] \setminus \{\node \}} & \\
    (\mathds{1}(\playertwo \in \subsetplayers) \numwin[\playertwo] + \mathds{1}(\playertwo \notin \subsetplayers) \numlose[\playertwo]) \Bigr] & \text{otherwise}
\end{array}
\right.
\label{eq:numlose}
\end{align}

\noindent
As $\parent{\treeroot} = \emptyset$, the product term for the root $\treeroot$ simplifies to 1.

The time complexity of computing the number of winning and losing coalitions, $\numwin[\node]$ and $\numlose[\node]$, respectively, at node $\node$ is $O(\branchingfactor 2^{\branchingfactor})$.
Traversing the game's hierarchy to compute these values for all  nodes $\node \in \nodes$ is a necessary first pass in computing \embpi, and hence \bpi.
The total cost of this first pass is thus $O(\numplayers \depth \branchingfactor 2^\branchingfactor)$.

The second pass of the algorithm then operates much like \mbpi.
The only difference is that the work done at each node along the path to a voter is $O(\branchingfactor 2^{\branchingfactor})$, not just $O(2^{\branchingfactor})$.
As such, this second pass, for a single voter, takes time $O(\depth \branchingfactor 2^{\branchingfactor})$, and thus for all voters $\player \in \players$ takes time $O(\numplayers \depth \branchingfactor 2^\branchingfactor)$.

Summing over both passes of the algorithm, the total complexity of computing \bpi{} via \embpi{} is $O(\numplayers \depth \branchingfactor 2^\branchingfactor)$.
Note that computing \bpi{} via \embpi{} for all voters is no more complex than computing \bpi{} for a single voter.
\section{Experiments: Vocabulary Selection}

Language has sentiment, typically either positive, negative, or neutral.
Furthermore, individual words have ``power'' in determining the sentiment of text.
For example, a review that reads ``The food is delicious.'' conveys a positive sentiment, stemming from the word ``delicious.''
We can thus view sentiment analysis as a voting game in which each of the words in some piece of text is a voter contributing to its overall sentiment.

The \textbf{vocabulary selection} problem is that of choosing a set of important words as a vocabulary for a natural language processing model \cite{patel2021}.
A smaller vocabulary makes models more interpretable \cite{adidi2018,samek2017}, requires less memory \cite{Sennrich2015}, is more amenable to use in a resource-constrained setting \cite{shi2017}, and is less prone to over fitting \cite{Chen2019,lhostis2016}.
\bpi{} has been used as a heuristic to solve the vocabulary selection problem, taking the power of individual words as a proxy for their importance \cite{patel2021}.

Computing \bpi{} for text can be costly, as sentences can be dozens of words long.
But language is compositional: individual words have meaning, which can be combined into phrases whose meaning derives from the individual words, which can be further combined into sentences, whose meaning derives from these phrases, and so on \cite{cogswell2019}.
Correspondingly, sentiments are typically compositional.
It is thus natural to model text sentiments, which derive from constituent words, as a hierarchical voting game.
Computing \bpi{} under this assumption 
can allow for an exponential speed up in computation (depending on the degree to which language is compositional).

The usual \mbpi{} formula, however, does not apply, as this voting game is not balanced.
If a collection of words in a text produces one sentiment, the opposite collection of words need not produce the opposite sentiment. 
For example, ``The overcooked vegetables were not good.'' ``not good" produces a negative sentiment but ``the overcooked vegetables were'' also produces a negative sentiment, not a positive sentiment.
Nonetheless, our \embpi{} formula, which applies to unbalanced hierarchical voting games, is applicable.

In this section, we describe experiments in which we compute the power of individual words in determining the sentiment of sentences extracted from customer reviews.
We use the Stanza library \cite{Qi2020} to ascribe both sentiment and hierarchical structure to these sentences.
As a characteristic function, we use Stanza's sentiment analysis classifier, with an output of 1 indicating positive sentiment, and 0 indicating neutral or negative sentiment.
We create a parse tree of the sentences using Stanza's constituency parser.

As a sentiment classifier, Stanza is an inexact model; so it may incorrectly classify some parts of a sentence, as compared to the whole.
For example, it might erroneously classify the word ``pretty'' as positive, even when it appears in the context ``pretty good,'' where it might better be classified as neutral, if not negative.

But even if classifed correctly, the sentiments of components of a text may not combine as expected. 
For example, ``no one with half a brain would think this food is bad'' has a positive sentiment, but both ``no one with half a brain'' and ``would think this food is bad'' have negative sentiment.
Indeed, sentences are \emph{not\/} monotonic in \samy{general}{their clauses}.
Nonetheless, we report anecdotally that non-monotonic sentences appeared infrequently in our data.
On the contrary, most texts combined as expected, given the sentiments of their constituent parts.
For example, ``Very professional`` (positive) and ``and excellent service`` (positive), combined to form ``Very professional and excellent service`` (positive).

We thus take as our characteristic function Stanza's sentiment classifier, using it to calculate our baseline, namely (naive) \bpi{} via \Cref{eq:bpi}, for small sentences.
We then enhance this computation with Stanza's parser, computing \mbpi{} and \embpi.

In sum, while BPI has been demonstrated to be an effective model for vocabulary selection, it is intractable to compute for long sentences.
As a potential remedy, we explore the accuracy and computational savings of \mbpi{} and \embpi{} compared to standard \bpi{}.

\subsection{Results}

We ran all experiments through Google Colab Pro, using a premium GPU, 
using the 2022 Yelp academic business review dataset \cite{yelp2014}.
We computed \bpi{} for sentences with lengths up 15 words, and \embpi{} and \mbpi{} for sentences with lengths up to 40.
To assess the accuracy of our approach, we calculate the mean squared error between the results of each algorithm and the \bpi{} baseline.%
\footnote{MSE has been used as a measure of the error in estimating \bpi{} before \cite{saavedra2021}.}

\begin{figure}[!ht]
    \begin{subfigure}{.48\textwidth}
        \centering
        \includegraphics[width=25em]{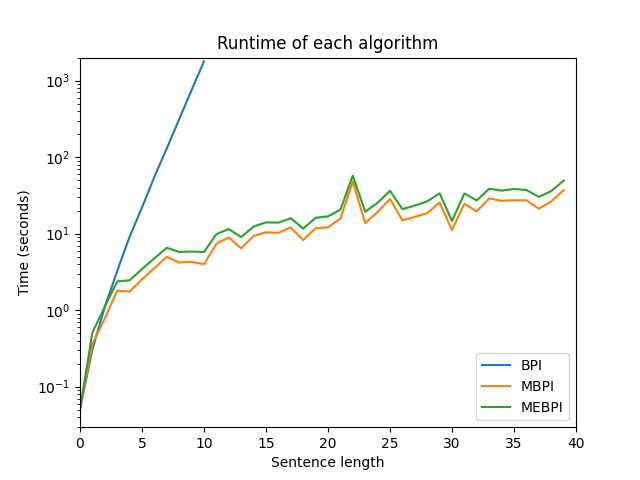}
        \caption{The run time to compute BPI, MBPI, and MEBPI as sentence length increases. BPI quickly becomes intractable, while MEBPI is on par with MBPI, despite its much broader applicability.}
        \label{fig:runtime}
        \end{subfigure}%
    
    \begin{subfigure}{.5\textwidth}
        \centering
        \includegraphics[width=25em]{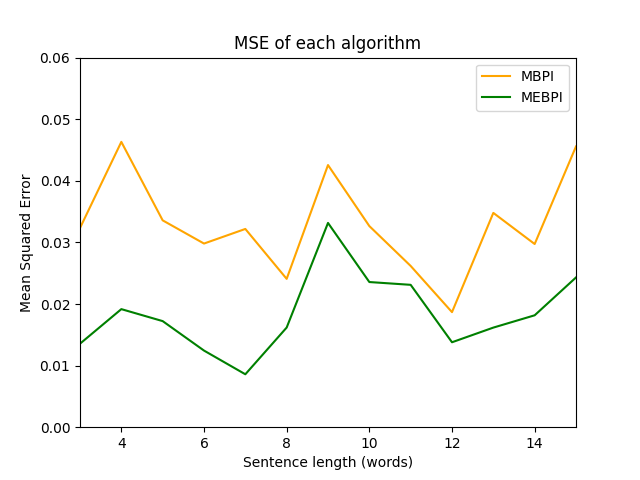}
        \caption{The MSE between MBPI \& BPI and between MEBPI \& BPI. MEBPI outperforms MBPI, in that it produces results that are more faithful to our baseline, BPI results.}
        \label{fig:MSE}
    \end{subfigure}
    \centering
    \caption{The runtime and error of computing MBPI and MEBPI as compared to BPI, when applied to the game of sentence sentiment.}
\end{figure}

We first report the algorithms' run times on sentences of different lengths (\Cref{fig:runtime}).
Then we report the mean squared error as the length of the sentence increases (\Cref{fig:MSE}).
Each data point is an average over 10 sentences of the specified length.

As this game is not balanced, as expected, we find that \embpi{} is more faithful to \bpi{} than \mbpi.
Moreover, as both rely on decompositions, they are both orders of magnitude faster than (naive) \bpi.
For example, on a sentence of length 10, (naive) \bpi{} took 700 seconds to run, while \mbpi{} took 4.3 seconds and \embpi{} took 5.8 seconds. 
On a sentence of length 11, (naive) \bpi{} took 1301 seconds to run, while \mbpi{} took 4.4 seconds and \embpi{} took 5.9 seconds.

\section{Conclusion}

This paper identified the key property that supports a natural decomposition of \bpi{} in hierarchical voting games, namely balanced.
We then introduced a novel power measure (\ebpi{}) that enabled us to calculate \bpi{} efficiently in games that are not balanced.
As an application domain, we studied the power of individual words in determining the sentiment of a sentence, taking advantage of the compositionality of language.
Decomposing the problem yielded order of magnitude speed ups, even for short sentences (10 words long), while maintaining relatively high accuracy.

There are many games that do not satisfy the definition of voting games studied here.
For example, in ternary voting games, the characteristic function is not binary-valued; voters can abstain, or words can be neutral in their sentiment \cite{felsenthal1997}.
Investigating whether results similar to ours apply in these games, and to alternative notions of power like SSPI, are interesting directions for future work.



\printbibliography


\end{document}